\documentclass[aps,prd,preprintnumbers,showpacs]{revtex4}
\usepackage[dvips]{graphicx}
\begin{document}

%
%

\eprint{Nisho-3-2019}
\title{Chiral Nonsymmetric Interaction in Strong Coupled QCD}
\author{Aiichi Iwazaki}
\affiliation{Nishogakusha University,\\ 
6-16 3-bantyo Chiyoda Tokyo 102-8336, Japan.}   
\date{Aug. 30, 2019}
\begin{abstract}
Chiral symmetry in massless QCD
is believed to be broken spontaneously.
We discuss a possibility that the chiral symmetry is explicitly broken by QCD monopoles which appear only in strong coupled QCD.
Namely, the monopole quark interaction explicitly breaks the chiral symmetry ( SU$_A(2)\times $U$_A$(1) ) just like bare quark mass terms.
We show that  
the strength of the interaction is roughly $10$ times smaller than standard strong interactions.
We describe it as an effective interaction $g'\bar{q}q\Phi^{\dagger}\Phi$ 
with the monopole field $\Phi$ and $g'$ being of the order of $(10\rm \,GeV)^{-1}$ or less.
It produces small constituent quark masses less than $1$MeV when the monopoles condense ( $\langle\Phi\rangle\neq 0 $ ).
We examine to what extent such a weak but explicit symmetry breaking interaction is allowed.
In particular, examining Gell-Mann-Oakes-Renner relation 
we find that the presence of such a small symmetry breaking term is still allowed within the present accuracy of lattice gauge simulations. 
We predict some phenomenological effects caused by the chiral nonsymmetric monopole quark interaction.
Quark confinement and chiral condensate ( $\langle\bar{q}q\rangle\neq 0$ ) 
arise simultaneously. The condensate $\langle\bar{q}q\rangle$ caused by the monopoles 
is proportional to monopole density and
is estimated such that $(-\langle\bar{q}q\rangle)^{1/3}\sim 160$MeV.   
The weak monopole quark interaction leads to
the small decay width of an observable monopole to hadrons.
\end{abstract}
\hspace*{0.3cm}

\hspace*{1cm}

\maketitle

\section{Introduction}
\label{1}
Massless QCD is classically chiral symmetric. It is widely believed that the symmetry holds
even quantum mechanically except U(1) chiral anomaly. At least, QCD is well defined perturbatively as chiral symmetric one.
That is, the chiral symmetric QCD can be defined in the weakly coupled region.
On the other hand, it is not obvious whether or not QCD is chiral symmetric in strong coupled region,
although the chiral symmetry is expected to be valid.
   
It is generally believed that pions are Nambu-Goldstone bosons in massless QCD.
Their small masses are caused by small bare quark masses. Namely,
the chiral symmetry of massless QCD is spontaneously broken in the strong coupled region
so that Nambu-Goldstone bosons appear; they are pions.
Based on the chiral symmetry in QCD, chiral perturbations\cite{chiral} have been formulated and
their validity has been examined in various ways.

In this paper, we discuss a possibility that massless QCD cannot be defined keeping the chiral symmetry
in the strong coupled region. The point is that QCD monopoles\cite{coleman} arising in the region
explicitly break the chiral symmetry; the monopole quark interaction inevitably breaks the chiral symmetry,
SU$_A(2)\times $U$_A(1)$. So, the chiral symmetry is not spontaneously but explicitly broken in the strong coupled QCD.
Thus, the pions are not Nambu-Goldstone bosons.
We discuss that the chiral symmetry breaking interaction is roughly $10$ times weaker than
the standard strong interactions between quarks and gluons. Because of its weak interaction, we may have missed it.

In order to see the possibility,
we examine how exactly the chiral symmetry is confirmed in lattice gauge theories.
In particular, we focus on the Gell-Mann-Oakes-Renner relation\cite{GMOR}.
The relation is an explicit result derived under the assumption of the chiral symmetry. 
It describes the relation among bare quark mass $m_{ud}$ and chiral condensate $\langle\bar{q}q\rangle$ as well as
observed pion mass $m_{\pi}$ and its decay constant $f_{\pi}$. 
Their values have been estimated independently in lattice gauge theories without using the relation.
The relation has been examined in detail and shown to hold very accurately.

However we find a possibility that small pion masses ( $0<m_{\pi}<60$MeV ) remain even in the chiral limit $m_{ud}\to 0$. 
The possibility arises mainly from the uncertainty in the evaluation of the chiral condensate $\langle\bar{q}q\rangle$. 
The strengths of the chiral nonsymmetric interactions are much weak compared with the chiral symmetric strong interactions.
The small pion mass ( $ <60$MeV ) remaining in the chiral limit is the result of the weak chiral nonsymmetric interaction.
( The interaction generates small quark masses $\sim 1$ MeV or less when the monopoles condense
so that the pions as bound states of the quarks acquire the small masses. )
Therefore, it appears that the presence of the chiral nonsymmetric weak monopole quark interaction cannot be excluded
at the present stage.

The presence of such an interaction also leads to an important consequence. That is,
the chiral condensate appears (  $\langle\bar{q}q\rangle\neq 0$ ) only when monopoles condense. 
That is, the chiral condensate takes place\cite{cc} simultaneously with quark confinement.

\vspace{0.2cm}
It is well known that monopoles in QCD play important roles for quark confinement\cite{dual}. 
The monopole condensation produces dual superconductor in which color electric fields
are squeezed into vortices.
The confinement picture 
has been extensively analyzed in lattice gauge theories. Especially, by using Maximal Abelian gauge\cite{mabelian},
the picture\cite{dual} of the confinement has been examined.

The monopoles have recently been discussed\cite{hasegawa} to play a role for chiral symmetry breaking.
In particular, their association with zero modes of Dirac operators has been examined in lattice gauge theory:
Addition of a pair of monopole and anti-monopole to gauge field configuration in Dirac operators
increases the number of the zero modes. 
It suggests that the monopoles play a role for the chiral symmetry breaking
as well as the quark confinement. It has also been discussed\cite{miyamura,suga,ciwazaki} that the monopole condensation
causes the chiral symmetry breaking.

Their observable effects have recently been analysed\cite{heavy} in the formation of quark gluon plasma ( QGP ).
In the analyses, the monopoles are quasi-particles in strong coupled QGP
and are a dominant component in the plasma.  
Furthermore, it has been discussed that 
they play a role of the decay of glasma produced just after high energy heavy ion collisions.
In this way, the roles of the monopoles in strong coupled QCD have been increasingly
recognized phenomenologically as well as theoretically.

In this paper, we analyse monopole quark interaction and find that it explicitly breaks the chiral symmetry.
The similar analysis has previously been performed to understand Rubakov effect\cite{rubakov,callan,ezawa}; baryon decay
in the scattering between baryon and GUT monopole. 
The previous analysis can be applied\cite{iwazaki} to the scattering between quark and QCD monopole. Then,
we find that chiral condensate  (  $\langle\bar{q}q\rangle\neq 0$ ) is locally present around the monopole\cite{ezawa}.
The condensate appears owing to the chiral nonsymmetric boundary condition for the quark at the location of  the monopole.
The condition explicitly breaks the chiral symmetry SU$_A(2)\times $U$_A(1)$.
Thus, the chiral condensate in vaccum receives the contribution of the monopoles.

We also discuss the generation of constituent quark masses $m_q$ by the monopole condensation.
It is caused by the boundary condition. The masses are found to be small such that $m_q$ is of the order of $1$MeV or less.
Then, the chiral condensate $\langle\bar{q}q\rangle$ in vacuum  additionally receives the contribution of the quarks
when the monopoles condense.

\vspace{0.1cm}
We discuss some of phenomenological effects of the chiral nonsymmetric monopole quark interactions.
Because the chiralities of quarks are not conserved in the presence of the monopoles,
chiral magnetic effects may not arise in high energy heavy ion collisions. 
We also predict that 
partial chiral restoration, i.e. the decrease of hadron masses, may arise in dense nuclear matters.
This is because the chiral condensate as well as monopole condensate
diminishes in the matters. Furthermore, we argue that the mixing between the monopole and isoscalar scalar meson $\bar{q}q$
is very small. Thus,
the decay width to hadrons is very small.

In the next section (\ref{2}), we discuss the monopole quark interactions and show that we need to impose a boundary condition
for quarks at the monopoles. The condition explicitly violates the chiral SU$_A(2)\times $U$_A$(1) symmetry. 
We show that the monopole
quark interaction is roughly $10$ times weaker than the other hadronic interactions.
In the section(\ref{3}), we show that the constituent quark mass $m_q$ is generated when the monopoles condense.
The mass is of the order of $1$MeV or less.
In the section(\ref{4}), we examine the accuracy of the Gell-Mann-Oakes-Renner relation
and find a possibility that the presence of the weak chiral symmetry breaking interaction is still allowed in the present accuracy of
the lattice gauge simulations. 
In the section(\ref{5}), we discuss the phenomenological effects of the chiral nonsymmetric monopole quark interaction.
Summary follows in the final section (\ref{6}).

\section{Chirality Nonconserved Monopole Quark Interaction}
\label{2} 
In order to explicitly treat the monopole quark interaction, we take SU(2) gauge theory with a quark doublet and
adopt the assumption of the Abelian dominance\cite{abelian,maximal}.
( The arguments in this section can be easily generalized\cite{iwazaki} to SU(3) gauge theory. )
It only holds in the low energy phenomena of QCD, that is, strong coupled QCD.
The validity of the assumption has been discussed in the lattice gauge theory by using Maximal Abelian gauge\cite{mabelian}.
According to the Abelian dominance, the relevant dynamical degrees of freedom are massless diagonal gluons, i.e. $A_{\mu}^{3}$, 
massless quarks $q^{\pm}$ of the doublet
and the monopole $\Phi$. The off-diagonal gluons are massive and irrelevant to the low energy phenomena. 
We suppose that gauge invariant quantities discussed in the paper are independent of
our choice of the Abelian dominance, although the validity of the Abelian dominance has been confirmed only with the Maximal Abelian gauge.

\subsection{Conservation of Angular Momentum under Magnetic field of Monopole}
First, we would like to show that the chiralities are not conserved in the monopole quark scattering. That is 
the chiral U$_A$(1) symmetry is broken.
According to the assumption of Abelian dominance, 
we consider the monopole quark interaction which is produced by the diagonal gauge field $A_{\mu}\equiv A_{\mu}^3$.
The quark doublet is composed of positive $q^+$ and negative $q^-$ ones associated with 
the charge $\pm g/2$ of the gauge field $A_{\mu}$.

Thus, we consider a massless quark doublet $\Big(\begin{array}{l}q^+ \\ q^-\end{array} \Big)$ scattered by the monopole, 

\begin{equation}
\label{1e}
\gamma_{\mu}(i\partial^{\mu}\mp \frac{g}{2}A^{\mu})q^{\pm}=0,
\end{equation}
where the gauge potentials $A^{\mu}$ denotes a Dirac monopole
given by 

\begin{equation}
\label{2e}
A_{\phi}=g_m(1-\cos(\theta)), \quad A_0=A_r=A_{\theta}=0
\end{equation} 
where $\vec{A}\cdot d\vec{x}=A_rdr+A_{\theta}d\theta+A_{\phi}d\phi$ with polar coordinates $r,\theta$ and $\phi=\arctan(y/x)$.
$g_m$ denotes a magnetic charge with which magnetic field is given by $\vec{B}=g_m\vec{r}/r^3$.
The magnetic charge satisfies the Dirac quantization condition $g_mg=n/2$ with integer $n$ where $g$ denotes the gauge coupling of SU(2) gauge theory.
Hereafter, we assume the monopole with the magnetic charge $g_m=1/2g$.

\vspace{0.1cm}

The monopole quark ( in general, fermion ) dynamics has been extensively explored, in particular, in the 
monopole catalysis\cite{rubakov,callan,ezawa} of baryon decay ( so called Rubakov effect ).
The important point we should stress is that conserved angular momentum has an additional component\cite{coleman}. That is,
it is given by

\begin{equation}
\label{3e}
\vec{J}=\vec{L}+\vec{S}\mp gg_m\vec{r}/r
\end{equation}
where $\vec{L}$ ( $\vec{S}$ ) denotes orbital ( spin ) angular momentum of quark.
The additional term $\pm gg_m\vec{r}/r$ play an important role of chiral symmetry breaking.
Owing to the term we can show that either the charge or the chirality is not conserved in the monopole quark scattering.
When the chirality ( or helicity $\sim \vec{p}\cdot\vec{S}/|\vec{p}||\vec{S}|$ ) is conserved, the spin must flip $\vec{S}\to -\vec{S}$ after the scattering 
because the momentum flips after the scattering; $\vec{p}\to -\vec{p}$.
Then, the charge must flip $g\to -g$ because of the conservation of $\vec{J}\cdot\vec{r}$, i.e.
 $\Delta(\vec{J}\cdot\vec{r})=\Delta(\vec{S}\cdot\vec{r})+\Delta(gg_mr)=0$. ( $\Delta(Q)$ denotes the change of the value $Q$ after the scattering. ) 
On the other hand, when the charge is conserved ( it leads to $0=\Delta(\vec{J}\cdot\vec{r})=\Delta(\vec{S}\cdot\vec{r}$) ),
the chirality $\vec{p}\cdot\vec{S}/|\vec{p}||\vec{S}|$ must flip because the spin does not flip $\vec{S}\to \vec{S}$.
Thus we find that either the charge or the chirality conservation is lost in the scattering.
We need to examine actual solutions in the scattering in order to determine which one is conserved.
( In the discussion we assume that the mass of the monopole is sufficiently large such that the collision of the quark
does not change the state of the monopole. That is, the energy of the quark is much less than the mass of the monopole. )

\subsection{Solutions of Quark Scattering with Monopole}
By solving the equation we find that an appropriate boundary condition is needed 
in order to define the quark scattering with monopoles. The boundary condition breaks either the charge conservation
or chirality conservation.
  
The components with $\vec{J}=\vec{L}=0$ are given by\cite{ezawa}

\begin{equation}
\label{4e}
q^{\pm}=\frac{1}{r}\left(\begin{array}{l}f_{\pm} (r,t) \\ \mp ig_{\pm}(r,t)\end{array}\right )\eta_{\pm} \quad \mbox{with} \quad
\frac{\sigma_i x_i}{r}\eta_{\pm}=\pm \eta_{\pm}.
\end{equation}
with Pauli matrices $\sigma_a$.

Then the equation is decomposed into two independent equations,

\begin{equation}
\label{5e}
i\bar{\gamma}_{\nu}\partial^{\nu}\psi_{\pm}=0 \quad \mbox{with} \quad \psi_{\pm}\equiv\left(\begin{array}{l}f_{\pm} (r,t) \\  -ig_{\pm}(r,t)\end{array}\right) 
\end{equation}
with $\nu=0,1$, $x_0=t$ and $x_1=r$, where two dimensional gamma matrices are defined by

\begin{equation}
\label{6e}
\bar{\gamma}^0=\left(\begin{array}{rr} 1 & 0 \\ 0  &-1 \\ \end{array}\right), \quad \bar{\gamma}^1=\left(\begin{array}{rr} 0 & 1 \\ -1  & 0 \\ \end{array}\right).
\end{equation}
We can easily solve the two dimensional equations. The solutions are characterized by their chiralities and charges as well as  
their motions i.e. incoming or outgoing.

When $E>0$, the solution $\psi_{+,R} =\exp(-iE(t-r))\left(\begin{array}{l}1 \\ 1\end{array}\right )$
 describes outgoing positive charge and right handed particles, while the solution  $\psi_{-,L} =\exp(-iE(t-r))\left(\begin{array}{l}1 \\ 1\end{array}\right )$
describes outgoing negative charge and left handed particles. Similarly, the solution $\psi_{+,L} =\exp(-iE(t+r))\left(\begin{array}{l}1 \\ -1\end{array}\right )$
describes incoming positive charge and left handed particles, while the solution  $\psi_{-,R} =\exp(-iE(t+r))\left(\begin{array}{l}1 \\ -1\end{array}\right )$
does incoming negative charge and right handed particles.

We note that the right ( left ) handed projection operator for $\psi_{+}$ is given  such that 
$\frac{1}{2}(1+\bar{\gamma}_5)$ ( $\frac{1}{2}(1-\bar{\gamma}_5)$ ) for $\psi_{+}$, while they are given such as 
$\frac{1}{2}(1-\bar{\gamma}_5)$ (  $\frac{1}{2}(1+\bar{\gamma}_5) $ )  for $\psi_{-}$
with $\bar{\gamma}_5=\left(\begin{array}{rr} 0 & 1 \\ 1  &0 \\ \end{array}\right)$.

There are four types of the solutions; incoming ones $\psi_{+,L}$ and $\psi_{-,R}$, while outgoing ones $\psi_{+,R}$ and $\psi_{-,L}$.
Obviously,  either of the charge or the chirality is not conserved. Therefore,
in order to define the scattering, we need to impose boundary conditions\cite{kazama} for the quarks at the location of the monopole.

\vspace{0.1cm}
The charge is strictly conserved because the charge conservation is guaranteed by the gauge symmetry.
When the quark flips its charge, monopoles are charged or heavy charged off-diagonal gluons must be produced to preserve the charge conservation.
But the processes cannot arise in the low energy scattering. Charged monopoles are dyons and they are heavy.   	
Thus, inevitably the chirality is not conserved. For instance, the right handed quark $q_R$ is transformed to the left handed quark $q_L$ in the scattering.
That is, 
we impose a boundary conditions $q^{\pm}_R(r=0)=q^{\pm}_L(r=0)$ at the location of the monopole, which breaks the chiral U$_A$(1) symmetry.

\subsection{Chiral NonConserved Boundary Condition}
The chiral symmetry breaking\cite{ciwazaki} has been rigorously shown\cite{rubakov,callan, ezawa} to be caused by chiral anomaly in QCD.
Even if we impose chirality conserved but charge nonconserved boundary conditions $q^+_{R,L}(r=0)=q^-_{R,L}(r=0)$ at the location of the monopole, 
we can show\cite{rubakov,callan,ezawa} that the charge is conserved,
but chirality is not conserved in the monopole quark scattering. 
The chirality nonconservation arises from chiral condensate $\langle\bar{q}q\rangle\propto 1/r^3$
locally present around the monopole at $r=0$.
The condensate is formed by the chiral anomaly when we take into account 
quantum effects of gauge fields $A_{\mu}=\delta A_{\mu}^{quantum}+A_{\mu}^{monopole}$.  
The condensate also arises when we choose the chirality nonconserved boundary condition.
Eventually, the chirality nonconserved boundary condition is realized in physical processes. 
These results are by-products in the analysis of the Rubakov effect\cite{rubakov,callan, ezawa}. 

Furthermore, it apparently seems that
quarks may change their flavors in the scattering. For instance, u quark is transformed into d quark.
Then, the monopole must have a SU(2) flavor after the scattering. But it is impossible because there are no such monopoles
with SU(2) flavors in QCD. They are flavor singlet.
Therefore, quarks cannot change their flavors in the scattering with the monopole.
Quarks change only their chiralities. It results in SU$_A$(2) chiral symmetry breaking as well as U$_{A}$(1) symmetry breaking.
Mathematically, the chiral nonsymmetric condensations arise such that  $\langle\bar{q}_iq_i\rangle\propto 1/r^3$ for each flavor $i$.
It will turn out in later section that the local chiral condensate around a monopole leads to the global chiral condensate
when the monopoles condensate in vacuum.

\vspace{0.1cm}
The chiral symmetry breaking arises at the location of the monopole. It is associated with properties at short distances.
So it is beyond the assumption of Abelian dominance.
But, the explicit symmetry breaking is not artifact of the assumption.
In QCD we have monopole solutions such as Wu-Yang monopoles. When we admit the presence of the monopoles
in QCD, we need appropriate boundary conditions\cite{kazama} for quarks at the locations of the monopoles
in order to quantum mechanically define the monopole quark scattering. The relevant boundary conditions are those such as boundary conditions 
breaking the chiral SU$_A$(2)$\times$U$_A$(1) symmetry:
$q^{i,\pm}_{R}(r=0)=q^{i,\pm}_L(r=0)$ where $i$ denotes a flavor. 
We may effectively describe the monopole quark interaction such that $\bar{q}^iq^i\Phi^{\dagger}\Phi$ with monopole field $\Phi$.
The interaction implies that quarks change their chiralities at the location of the monopole.  

%

\subsection{Weak Monopole Quark Interaction}
Now, we show that the monopole quark interaction is much weak compared with the ordinary strong interactions whose
strengths are controlled by $\alpha_s=g^2/4\pi$.
The quarks have color charges $g/2$ and the monopoles have magnetic charges $g_m=1/2g$.
Therefore, the monopole quark interaction ( $\propto g\times g_m$ ) 
does not involve $\alpha_s$ at tree level. On the other hands, the quark gluon interactions involve
$\alpha_s$ even at tree level.  
It implies that the monopole quark interaction is weak compared with the quark gluon interactions. 
The ratio of the strength of the monopole quark interaction to those of the quark gluon 
interactions is naively of the order of $(g/2\times g_m)/g^2=1/16\pi\alpha_s\simeq 1/20$ with $\alpha_s(1\rm GeV)\simeq 0.5$.  
Therefore, 
the chiral nonsymmetric interaction is roughly $10$ times weaker than the ordinary strong interactions.
The chiral SU$_A$(2) symmetry approximately holds. 
Because of its weak interaction, we may have missed it.

The weak monopole quark interaction leads to the small mixing of the monopole with isoscalar scalar $\bar{q}q$.
The monopole is a glueball. In general, it appears that glueballs can easily mix with the quarks $\bar{q}q$.
So it is difficult to distinguish glueballs from isoscalar mesons. But the monopole can be easily
distinguished from the mesons, because it hardly mixes with the isoscalar mesons.

\vspace{0.1cm}
It is the standard idea that when we remove small bare quark mass terms $m_{ud}$,
the chiral symmetry holds and it is spontaneously broken.
Then, we may regard the pions as Nambu-Goldstone bosons.
Similarly, we expect that the symmetry is spontaneously broken
when we remove the chiral nonsymmetric monopole quark interaction.
But it is impossible because there are no free parameters to adjust the strength of the interaction.
( In the case of the quark mass terms, we have adjustable parameters, that is, bare quark masses $m_{ud}$. ) 
Thus, the chiral symmetry is not spontaneously broken; the chiral nonsymmetric interaction cannot be removed by hand.
The pions are not Nambu-Goldstone bosons.

\section{Quark Mass Generation by Monopole Condensation}
\label{3}

We discuss that quark masses are generated when the monopoles condense in massless QCD.
The generation of the quark masses leads to the chiral condensate $\langle\bar{q}q\rangle\neq 0$.
The masses are expected to be small because the monopole quark interaction is weak.

Quarks change their chirality without the change of their flavors at the location of the monopole.
We can describe such a interaction by using monopole field $\Phi$,

\begin{equation}
\label{7e}
g'|\Phi|^2(\bar{u}_Lu_R+\bar{u}_Ru_L+\bar{d}_Ld_R+\bar{d}_Rd_L)=g'|\Phi|^2(\bar{u}u+\bar{d}d).
\end{equation}

The parameter $g'$ is of the order of the inverse of $\Lambda_{QCD}$ times $gg_m/(2g^2)=1/16\pi\alpha_s$, that is, 
$g'\simeq \Lambda_{QCD}^{-1}/\big(16\pi\alpha_s(\Lambda_{QCD})\big)$. ( Physically speaking, the mass dimension of $g'^{-1}$
is determined by the mass of the monopole proportional to $\Lambda_{QCD}$. )
Numerically, it is given such that $g'\sim (25.2\mbox{GeV})^{-1}$
with $\Lambda_{QCD}=1$GeV and $\alpha_s(Q=1\mbox{GeV})\simeq 0.5$. 
The estimation is very rough. When we take $g'\simeq (3\Lambda_{QCD})^{-1}/\big(16\pi\alpha_s(\Lambda_{QCD})\big)$,
then, it leads to $g'\simeq (75.6\rm GeV)^{-1}$.
Although the precise value of $g'$ is not clear,
it would take a value in the range of
$(10\mbox{GeV})^{-1} \sim (100\mbox{GeV})^{-1}$.

Obviously, the interaction in eq(\ref{7e}) generates the quark mass $m_q=g'v^2$ when the monopole condenses $v=\langle\Phi\rangle\neq 0$.
In order to estimate the mass, we use a value estimated in a dual superconducting model\cite{dual,suzuki,maedan} of quark confinement.
It is given\cite{che} by $\langle\Phi\rangle=175$MeV. 
Therefore, we obtain the quark mass $m_q$ being of the order of $1$ MeV or less ( but at least, it would be bigger than $0.1$MeV). 
It is comparable to the bare u and d quark masses.
The smallness comes from the weak monopole quark interaction.
We should note that the estimation of the mass $m_q$ should not be taken seriously because
it is very rough. But,
it is important to note that the small mass $m_q$ comes from the weak monopole quark interaction.

We should note that the mass $m_q$ generated by the monopole condensation is a constituent quark mass
and is not bare ( current ) quark mass. The mass $m_q=g'v^2$ varies depending on the environment in which
the quarks are present. This is because the monopole condensate $v=\langle\Phi\rangle$ depends on the
environment, e.g. inside of hadrons, dense nuclear matters, e.c.t..

We should mention that the chiral condensate breaks the chiral SU$_A$(2)$\times$U$_A$(1) symmetry.
It is explicitly broken by the monopole quark interaction in eq(\ref{7e}).
Thus, the pions are not Nambu-Goldstone bosons. We expect that their nonvanishing masses are small
because the constituent quark mass $m_q$ composing the pions are small. 
In the next section, we discuss to what extent the nonvanishing pion masses are allowed in the chiral limit by
examining the uncertainty of numerical simulations in lattice gauge theories.

\section{Gell-Mann-Oakes-Renner relation}
\label{4}

We examine to what extent the idea of the spontaneous chiral symmetry breaking is exact.
According to the idea, the pions as Nambu-Goldstone bosons must satisfy the Gell-Mann-Oakes-Renner ( GMOR ) relation\cite{GMOR},

\begin{equation}
\label{8e}
m^2_{\pi}f^2_{\pi}=(2m_q+ 2m_{ud})(-\langle\bar{q}q\rangle)
\end{equation}
with $m_{ud}=(m_u+m_d)/2$ and the pion decay constant $f_{\pi}\simeq 92$MeV,
where we put an extra term $2m_q(-\langle\bar{q}q\rangle)$.
The term may be present if the chiral symmetry is explicitly broken in the limit $m_{ud}\to 0$.

The formula receives small corrections\cite{review} due to the quark mass terms. Namely, 
\begin{eqnarray}
\label{9e}
m^2_{\pi}&\simeq& m^2\big(1-(m^2/32\pi^2f^2)\log(\Lambda^2/m^2)\big) \quad 
\mbox{with} \quad \frac{m_{\pi}^2}{32\pi^2f^2_{\pi}}\simeq 0.007 \\
f^2_{\pi}&\simeq& f^2\big(1+(m^2/16\pi^2f^2)\log(\Lambda^2/m^2)\big) \quad 
\mbox{with} \quad \frac{m^2_{\pi}}{16\pi^2f_{\pi}^2}\simeq 0.014
\end{eqnarray}
with $\Lambda$ being of the order of $\Lambda_{QCD}$, where
the parameters $m$ and $f$ are given by $m(m_{ud})=m_{\pi}$ and $f(m_{ud})=f_{\pi}$ when $m_{ud}=0$. Thus, $m_{\pi}=0$ when $m_{ud}=0$. 
The GMOR relation is given by $m^2f^2=2m_{ud}(-\langle\bar{q}q\rangle)$ in terms of the variables $m$ and $f$.

These
chiral corrections of the bare quark mass to the GMOR formula $m^2_{\pi}f^2_{\pi}=2m_{ud}(-\langle\bar{q}q\rangle)$
has been numerically estimated\cite{correction} such that $m^2_{\pi}f^2_{\pi}(1-\delta)=2m_{ud}(-\langle\bar{q}q\rangle)$ with $\delta\simeq 0.047\pm 0.017$.
The parameter $\delta$ represents the accuracy of the relation.
We examine to what extent the mass term $m_q$ is allowed in the simulations of lattice gauge theories.

Using Banks-Casher relation\cite{BC}, the chiral condensate has been obtained\cite{chiralcon} such that $(-\langle\bar{q}q\rangle)^{1/3}=270$MeV$\pm 5$MeV.
On the other hand, the quark mass has been estimated\cite{review} such that $m_{ud}=3.4$MeV$\pm 0.1$MeV.
Furthermore, the pion decay constant has been measured such that $f_{\pi^0}=92$MeV$\pm 3$MeV and $f_{\pi^{\pm}}=92$MeV$\pm 0.3$MeV.   

When we use the values of the chiral condensate $ -\langle\bar{q}q\rangle\simeq (270\rm MeV)^3\big(1\pm (15/270)\big)$
and the quark mass $m_{ud}\simeq 3.4\rm MeV\big(1\pm (0.1/3.4)\big)$, the GMOR relation is well satisfied within the uncertainty such that
$2m_{ud}(-\langle\bar{q}q\rangle)/\big((1-\delta)m_{\pi}^2f_{\pi}^2\big)\simeq (0.93\sim 0.89)(1\pm 0.09)=1.01\sim 0.81$
with the use of $\delta=0.065\sim 0.03$, $m_{\pi}=135$MeV and $f_{\pi}=92$MeV.

But the ambiguities are no so small for the extra quark mass $0<m_q< 0.8$MeV to be excluded; 
$0.2\simeq -2m_q\langle\bar{q}q\rangle/(m_{\pi}^2f_{\pi}^2)$ with $m_q=0.8$MeV and $-\langle\bar{q}q\rangle=(270\rm MeV)^3$ .
It allows the presence of the nonzero
pion mass $0<m_{\pi 0}<60$MeV in the chiral limit.
(  The mass of the pion is given by $m_{\pi0}=\sqrt{(2m_q)(-\langle\bar{q}q\rangle)}/f_{\pi}$ in the chiral limit $m_{ud}=0$. )
The allowed value of the quark mass $m_q$ is consistent with our rough estimation in the previous section.  

\vspace{0.1cm}
Therefore,
the idea that the chiral symmetry is explicitly, but weakly broken even in the chiral limit is not excluded. The pions would have
nonzero masses in the limit. Their masses would be of the order of $10$MeV or less because the constituent quark mass
takes a value ( $0.8$MeV $>m_q> 0.3$MeV ); the lower limit on $m_q$ is obtained from
the strength of the monopole quark interaction $g'=(100\mbox{GeV})^{-1}$.

\section{Phenomenological Effects of Explicit Chiral Symmetry Breaking}
\label{5}

We would like to point out phenomenological consequences 
arising from the chiral nonsymmetric monopole quark interaction.

\subsection{Absence of Chiral Magnetic Effect}
First,
the monopole quark interaction gives rise to an important effect 
in strong coupled QGP produced by high energy heavy ion collisions. 
It has recently been pointed out that chiral magnetic effects\cite{chiralmagnetic} can be seen in the plasma.
The effects are caused by chiral imbalance  (the number of quarks with positive chirality
minus the number of quarks with negative chirality )
in the plasma. The imbalance is produced in the early stage of the high energy heavy ion collisions. It is the stage of glasma decay. 
The imbalance is represented by chiral chemical potential $\mu_5$. 
With the nonzero chiral chemical potential $\mu_5\neq 0$, electric currents are induced parallel to magnetic fields produced in the ionized heavy ion collisions.
The magnitudes of the currents are proportional to $\mu_5$.
It is called as chiral magnetic effect. At the temperatures near the crossover in quark-hadron phase transition,
the monopoles have been discussed\cite{heavy} to be predominant quasiparticles in the QGP and to interact with quarks at the temperatures.
Then, the chiralities of the quarks are not conserved. Thus, 
the chiral imbalance $\mu_5\neq 0$ is washed out by the monopoles.
That is, $\mu_5=0$ in the QGP.
Therefore, the chiral magnetic effects cannot be observed; the chiral imbalance vanishes in the monopole dominant phase of the QGP. 

\subsection{Chiral Condensate and Monopoles }
Secondly, owing to the monopole quark interaction $g'|\Phi|^2\bar{q}q$, 
the chiral condensate $\langle\bar{q}q\rangle\neq 0$ arises only 
when the monopoles condense $\langle\Phi\rangle\neq 0$.
The condensation leads to the quark mass which causes the nonvanishing chiral condensate.
Therefore, it is not accidental that the quark confinement and the chiral condensate appear
at an identical temperature\cite{cc}. Furthermore, we can show that the contribution of the monopoles is given such as
the chiral condensate $\langle\bar{q}q\rangle$
being proportional to the monopole density. 
The feature comes from the fact that the chiral condensate locally arises around each monopole.
That is, each monopole has been shown to carry the local chiral condensates\cite{ezawa},

\begin{equation}
\overline{\langle\bar{q}q\rangle}_m\equiv\int \frac{d\Omega}{4\pi}\langle\bar{q}q\rangle_m=-\frac{1}{(4\pi)^2r^3},
\end{equation}
where we have taken the average over the angles $\theta$ and $\phi$, $\int d\Omega\, \eta^{\dagger}(\theta,\phi)\eta(\theta,\phi)$=1.
( In the argument, the components $\vec{J}=\vec{L}=0$  of solutions in eq(\ref{4e}) have been analyzed\cite{ezawa}. Thus, $q\propto \eta(\theta, \phi)$. )
The formula holds irrespective of the sign of magnetic charges, that is, positive or negative one.

When two monopoles are located at $\vec{r}=\vec{r}_1$ and $\vec{r}=\vec{r}_2$, we may have

\begin{equation}
\label{12e}
\overline{\langle\bar{q}q\rangle}_m=-\frac{1}{(4\pi)^2|\vec{r}-\vec{r}_1|^3}-\frac{1}{(4\pi)^2|\vec{r}-\vec{r}_2|^3}
\end{equation}
The formura holds only for $|\vec{r}_1-\vec{r}_2|\gg |\vec{r}-\vec{r}_{1,2}|$. Thus,
supposing that the spacially homogeneous number density of the monopoles are given by $n_m\equiv n_m^{+}+n_m^{-}$,
we obtain

\begin{equation}
\label{13e}
\overline{\langle\bar{q}q\rangle}_m=-\int d^3x \frac{n_m(\vec{x})}{(4\pi)^2|\vec{r}-\vec{x}|^3}=-\frac{n_m}{4\pi}\log(l/r_c),
\end{equation}
with the average length $l=n_m^{-3}$ between monopoles and the minimum length $r_c\sim \Lambda_c^{-1}$ for the Abelian dominance to hold.
( The superposition of the chiral condensate around each monopoles may be allowed only for
local region $\vec{x}$ near a monopole at $\vec{r}$, which is separated by the distance $l$ from the other monopoles; $l\gg |\vec{x}-\vec{r}|$.) 

The number density $n_m$ behaves smoothly in the following. That is,  it starts at zero in high temperature and increases as the temperature decreases.
Then, it passes the value at the transition temperature between confinement and de-confinement. Finally, it reaches 
the maximum value as a condensed state of the monopoles in lower temperature.
Thus, the chiral condensate $\langle\bar{q}q\rangle_m\propto n_m$
also increases from  $\langle\bar{q}q\rangle_m=0$ and reaches the maximum value as the temperature decreases. 
Therefore, the chiral condensate does not show the sharp transition between $\langle\bar{q}q\rangle_m=0$ in high temperature
and $\langle\bar{q}q\rangle_m\neq 0$ in vacuum.

\vspace{0.1cm}
The number density $n_m$ of the monopoles is not equal to the magnetic charge density. 
Field theoretically, the number $N=\int d^3x\,\, n_m(\vec{x})$ of the monopoles is represented in the following,

\begin{eqnarray}
N&=&\int d^3k (a_+^{\dagger}(\vec{k})a_+(\vec{k})+a_-^{\dagger}(\vec{k})a_-(\vec{k})) \nonumber \\
&=&
\int d^3xd^3y(\Delta_1(\vec{x}-\vec{y})\partial_t\Phi^{\dagger}(\vec{x},t)\partial_t\Phi(\vec{y},t)+\Delta_2(\vec{x}-\vec{y})\Phi^{\dagger}(\vec{x},t)\Phi(\vec{y},t),
\end{eqnarray}
with creation ( annihilation ) operators $a_{\pm}^{\dagger}(\vec{k})$ ( $a_{\pm}(\vec{k})$ ) of positive or negative magnetic charged monopoles,
where 

\begin{equation}
\Delta_1(\vec{x})\equiv\frac{\int d^3k \exp(i\vec{k}\vec{x})}{2(2\pi)^3\sqrt{\vec{k}^2+M^2}}, \quad 
\Delta_2(\vec{x})\equiv\frac{\int d^3k \sqrt{\vec{k}^2+M^2}\exp(i\vec{k}\vec{x})}{2(2\pi)^3}.
\end{equation}
with the mass $M$ of the monopoles. We have neglected the contribution of dual gauge fields.

We estimate the chiral condensate using the above formula eq(\ref{13e}).
It is easy to see that the number density of the condensed monopoles is given by $n_m=M\langle\Phi\rangle^2$. 
Assuming $r_c=(\Lambda_c)^{-1}=(1\rm GeV)^{-1}$, 
we obtain the value $(-\overline{\langle\bar{q}q\rangle}_m)^{1/3}=(M \langle\Phi\rangle^2/4\pi\times \log(n_m^{-1/3}/r_c))^{1/3}\simeq 155$MeV
with $\langle\Phi\rangle=175$MeV and $M=1500$MeV, where we have identified the monopole as $f_0(1500)$ meson. ( See subsection (\ref{5.4}). )

\vspace{0.1cm}
The value $(-\langle\bar{q}q\rangle_m)^{1/3}\simeq 155$MeV contains only the contribution of the monopoles. 
When the monopoles condense, the quarks acquire the mass $m_q$ so that
the contribution of the quarks is added to the value $155$MeV. 
In the real situation, we have the bare quark mass $m_{ud}$ in addition to $m_d$.
Thus, using the free fields of the quarks, we obtain the contibution of the quarks to the chiral condensate,

\begin{equation}
\langle\bar{q}q\rangle_q=-\frac{3(m_q+m_{ud})^3}{\pi^2}\big(x\sqrt{x^2+1}-\log(x+\sqrt{x^2+1})\big),
\end{equation}
with $x=\Lambda_c/(m_q+m_{ud})$,
where we have taken into account the degrees of freedom of $3$ colors and $2$ flavors.
It leads to $(-\langle\bar{q}q\rangle_q)^{1/3}\simeq 107$MeV
with the use of the values $m_q+m_{ud}=4$MeV and $\Lambda_c=1$GeV; 
the cutoff $\Lambda_c$ is the momentum scale beyond which the Abelian
dominance does not hold.
Therefore, adding both contributions of the monopoles and quarks, 
we obtain $(-\langle\bar{q}{q}\rangle)^{1/3}\equiv (-\langle\bar{q}{q}\rangle_m-\langle\bar{q}{q}\rangle_q)^{1/3}\simeq 175$MeV.
Our estimation is very rough. But the value $175$MeV is not so far from the estimation $270$MeV in lattice gauge theories.

\subsection{Decrease of Hadron Mass in Dense Nuclear Matter}
Thirdly, 
the masses of hadrons may decrease in dense nuclear matters. This is because the chiral condensate decreases
as the monopole condensate does in the dense nuclear matters; $\overline{\langle\bar{q}q\rangle}\propto\langle\Phi\rangle^2$. 
The masses of the hadrons depends on
the chiral condensate just like the pion mass in eq(\ref{8e}). 
Then, as the monopole condensate decreases, the masses also decrease. 

It is easy to see the effect of the dense nuclear matter on the chiral condensate or monopole condensate.
Since the chemical potential $\mu_q$ of the quark number is present in the Lagrangian $\mu_qq^{\dagger}q$, the one loop effect of the quark $q$
on the ground state energy is represented by

\begin{eqnarray}
-2N_cN_f\mu_q\int \frac{d^3p}{(2\pi)^3}\theta(p_f-|\vec{p}|)&=&-\frac{4(\mu_q^2-m_q^2)^{3/2}\mu_q}{3\pi^2}
   \nonumber \\
 &\simeq& -\frac{4\mu_q^4}{3\pi^2}+\frac{2\mu_q^2m_q^2}{\pi^2}  \quad \mbox{with} \quad p_f=\sqrt{\mu_q^2-m_q^2},
\end{eqnarray}
with the number of color $N_c=2$ ( flavor $N_f=2$ ),
where $m_q=g'v^2$ denotes the quark mass generated by the monopole condensate $v=\langle\Phi\rangle$.
We have assumed $\mu_q \gg m_q$ because the mass $m_q$ is much small ( $m_q<1$MeV).
Obviously, this potential energy ( $\propto \mu_q^2m_q^2=+g'^2\mu_q^2v^4$ ) for $v=\langle\Phi\rangle$
makes the monopole condensate $v$ decrease as the chemical potential $\mu_q$ increases. 
Therefore, the masses of the hadrons decrease in the dense nuclear matter.

In the denser nuclear matters
than the average nuclear density, the fluctuations of color electric fields are larger than those in the nuclei. The large fluctuations
of the color electric fields make the monopole condensate decrease.
This is because the color electric fields expel the monopole condensate.
Therefore, we expect that the masses of the hadrons decrease
in the dense nuclear matters. The argument is coincident with the above analysis.

\subsection{Small Decay Width of Monopole}
\label{5.4}
Finally, the decay width of physical monopole into hadrons is much small compared with those of hadrons. 
The monopole is a glueball and it decays producing light quarks. But the coupling with the quarks is roughly $10$ times weaker than 
the standard strong interactions. Thus,
the hadronic decay width is expected to be small. 
The weak monopole quark interaction also leads to the small mixing between the monopole and an isoscalar scalar meson $\bar{q}q$.
This small mixing has been shown\cite{mix} in lattice gauge theories. 

As we have mentioned in the above section, in general, glueballs can mix with isoscalar scalar mesons
because gluons strongly couple with quarks. Such glueballs are composed of off-diagonal gluons in the picture of Abelian dominance.
They strongly couple with quarks. 
But the monopole as a glueball only couple with quarks very weakly. 
Furthermore,
the monopole couplings with off-diagonal gluons are much small because the off-diagonal gluons have color charge $g$ just like quarks.
Hence, the monopole hardly mixes with the isoscalar mesons through the off-diagonal gluons as well as quarks.

In addition, the monopole does not directly couple with photons.
It decays into photons only through the coupling with quarks. Thus, two photon decay width is much smaller than
the small hadronic decay width.

Therefore, we predict that the natural candidate of the monopole is the meson $f_0(1500)$.

\section{Summary}
\label{6}

We have analyzed the monopole quark interaction in strong coupled region of QCD.
We have found that it explicitly breaks the chiral symmetry SU$_A$(2)$\times$U$_A$(1).
It's strength is 10 times weaker than those of standard strong interactions controlled by $\alpha_s$. 
Thus, it is difficult to find the existence of such an interaction. 
It would lead to the general consensus that the chiral symmetry is spontaneously broken. 

In order to see the existence of such an explicitly chiral symmetry breaking interaction,
we have examined how exactly the Gell-Mann-Oakes-Renner relation is confirmed in
lattice gauge theories. We have found that a small quark mass $m_q$ generated by the interaction is allowed
as long as $m_q<0.8$MeV. It leads to the nonvanishing pion masses $m_{\pi}<60$MeV even in the chiral limit.

The monopole quark interaction also induces the chiral condensate $\langle\bar{q}q\rangle$ proportional
to the monopole density $n_m$ in the chiral limit. We have estimated the value $(-\langle\bar{q}q\rangle_m)^{1/3}\sim 160$MeV
using $\langle\Phi\rangle=175$MeV in a dual superconducting model and
assuming that the monopole is identified as the isoscalar scalar meson $f_0(1500)$.
As well as the monopoles, the quarks also contribute to the condensate, when
they acquire their masses. The contribution of the quarks arises in the chiral limit only when the monopole condensation takes place in the vacuum. 
Both contributions would lead to 
the value $(-\langle\bar{q}q\rangle)^{1/3}\sim 270$MeV
obtained in lattice simulations.

We have also discussed that hadron masses decrease in dense nuclear matter because the monopole
density $n_m\propto \langle\Phi\rangle^2$ decreases in the matter; the masses depend on the chiral condensate which is proportional to $n_m$.  

As has been shown in lattice gauge theories, the mixing between a glueball with the minimum mass and the scalar meson $\bar{q}q$ is very small.
The fact can be understood in our analysis because the monopole as a glueball hardly mixes with the meson $\bar{q}q$.
It comes from the weak monopole quark interaction as well as weak monopole off-diagonal gluon interactions.
Hence, we identify the monopole with the meson $f_0(1500)$ whose decay width is 
relatively small.

\section*{Acknowledgment}

The author
expresses thanks to Dr. H. Fukaya, Osaka University, Dr. M. Hasegawa, BLTP Joint Institute for Nuclear Reseach
and Prof. O. Morimatsu, KEK for useful comments
and discussions. This work is supported by in part by Grant-in-Aid for Scientific Research ( KAKENHI ), No.19K03832.

\end{document}